\documentclass{emulateapj}
\usepackage[utf8]{inputenc}
\usepackage{natbib}
\usepackage{graphicx}
\usepackage{multirow} 

\usepackage[colorlinks,linkcolor=blue,anchorcolor=blue,citecolor=blue]{hyperref}

\begin{document}
\title{Identification of galaxy cluster substructures with the Caustic method}

\author{Heng Yu\altaffilmark{1,2,3}, Ana Laura Serra\altaffilmark{4}, Antonaldo Diaferio\altaffilmark{1,3}, 
Marco Baldi\altaffilmark{5,6,7}}

\altaffiltext{1}{Dipartimento di Fisica, Universit\`a di Torino, Via
  P. Giuria 1, I-10125 Torino, Italy}
\altaffiltext{2}{Department of Astronomy, Beijing Normal University,
    Beijing 100875, China }
\altaffiltext{3}{Istituto Nazionale di Fisica Nucleare (INFN), Sezione di Torino, Via
  P. Giuria 1, I-10125 Torino, Italy}
\altaffiltext{4}{Dipartimento di Fisica, Universit\`a di Milano, Via Celoria 16, 20133 Milano, Italy}
\altaffiltext{5}{Dipartimento di Fisica e Astronomia, Alma Mater Studiorum Universit\`{a} di 
                   Bologna, viale Berti Pichat, 6/2, I-40127 Bologna, Italy}
\altaffiltext{6}{INAF - Osservatorio Astronomico di
                   Bologna, via Ranzani 1, I-40127, Bologna, Italy}
\altaffiltext{7}{INFN - Sezione di Bologna, viale Berti Pichat 6/2, 
                   I-40127, Bologna, Italy}

\begin{abstract}
We investigate the power of the caustic technique for identifying substructures of galaxy clusters
from optical redshift data alone. The caustic technique is designed to estimate the mass profile of galaxy 
clusters to radii well beyond the virial radius, where dynamical equilibrium does not hold.
Two by-products of this technique are the identification of the cluster members and the identification of the cluster
substructures. We test the caustic technique as a substructure detector on 
two samples of 150 mock redshift surveys of clusters; the clusters 
are extracted from a large cosmological $N$-body simulation of a $\Lambda$CDM model and 
have masses of $M_{200} \sim 10^{14} h^{-1} M_{\odot}$
and $M_{200} \sim 10^{15} h^{-1} M_{\odot}$ in the two samples. We limit
our analysis to substructures identified in the simulation
with masses larger than $10^{13} h^{-1} M_{\odot}$.
With mock redshift surveys with 200 galaxies 
within $3R_{200}$, (1) the caustic technique recovers $\sim 30-50$\% of the real substructures, 
and (2) $\sim 15-20$\% of the substructures identified by the caustic technique correspond to 
real substructures of the central cluster, the remaining fraction being low-mass substructures, 
groups or substructures of clusters
in the surrounding region, or chance alignments of unrelated galaxies.  
These encouraging results show that the caustic technique is a promising
 approach for investigating the complex dynamics of galaxy clusters. 
\end{abstract}


\section{Introduction}
Galaxy clusters, as the high-mass tail of the hierarchical structure,
connect the large-scale structure to galaxies, and 
are thus relevant for constraining models of galaxy evolution, structure
formation, and cosmology. 
Due to the large scale and long relaxing time of clusters, 
the presence of substructures is quite common, indicating  
that a number of clusters are likely to be out of equilibrium.  
Substructures can substantially affect the estimate of the
cluster velocity dispersion and mass \citep{1996Girardi,1996Pinkney}, 
can provide insights into the formation process of the 
cluster, and unveil the
existence of dark matter \citep{2004Markevitch,2006Clowe}.
The presence of substructures appears to be a fundamental ingredient
of the galaxy-environment connection and for shaping the morphology-density
relation \citep[e.g.,][]{2015Fasano, 2015Girardi}.  
The mass fraction in substructures can also probe structure formation and  the
expansion rate of the universe \citep{1992Richstone,1993Kauffmann,1995Mohr,1998Thomas}. 

Attempts to identify and investigate cluster substructures have been numerous since their
first discovery in the optical band \citep{1954Shane}.
In images of X-ray surface brightness, substructures are relatively obvious,
especially with data coming from recent X-ray missions, including {\sl ROSAT}
\citep{2001Kolokotronis,2001Schuecker}, {\sl Chandra}\citep{2005Jeltema,2012Andrade-Santos,2014Parekh},
and {\sl XMM-Newton}\citep{2009Zhang}.

However, because the X-ray surface brightness decreases rapidly with increasing radius,
these observations can only trace substructures in the central region of clusters.
Substructures can also appear in microwave observations 
because free electrons in the hot X-ray gas originate the Sunyaev-Zeldovich effect 
\citep{2001Komatsu,2011Korngut}.
 The location of radio halos in 
clusters also tends to coincide with the substructures observed in X-ray images and
temperature maps \citep[see][for a recent review]{2012Feretti}. 

The existence of substructures 
in the dark matter halos of clusters can 
also be revealed by the anomalous images of strong gravitational lensing 
systems \citep{1996Kneib,1998Mao,2004Mao}, 
or by peculiar features of the halo density profiles of weak lensing systems
\citep{2000Hoekstra,2006Clowe,2010Okabe,2011Mira,2013Oguri,2015McCleary,2015Shirasaki}, 
although the contamination by chance alignments of unrelated massive
systems along the line of sight can be severe \citep{2003Hoekstra,2011Hoekstra,2013geller}. 

The detection of substructures from optical data, based on the
galaxy celestial coordinates and redshifts, is still a common approach for studying substructures. 
The methods can either use galaxy positions alone, redshifts alone, or both.

The methods that use galaxy positions alone include 
the smoothed density-contour maps \citep{1982Geller}, symmetry test, angular
separations test, density contrast test \citep{1988West}, average two-point correlation 
function \citep{1993Salvador}, and two-dimensional (2D) wavelet transforms
 \citep{1990Slezak,1995Escalera,2006Flin}. They
usually suffer from the contamination of background and foreground galaxies, 
but are clearly very useful when spectroscopic observations are missing or incomplete. 

The methods that only use the galaxies' redshifts usually assume that the distribution of
the velocities of the member galaxies is Gaussian. 
Based on this assumption, 
the indicators like kurtosis, skewness \citep{1990West, 1999Solanes}, 
and the asymmetry and tail indices \citep{1993Bird} are 
designed to quantify the subclustering in the one-dimensional (1D) redshift distribution.
The 1D Kaye's mixture model (KMM) algorithm belongs to the set of Gaussian mixture model (GMM) methods:
it assesses the presence of substructures
by estimating the number of optimal partitions of Gaussian distributions
\citep{1994Ashman, 1997Kriessler}.
The DEDICA method is based on an adaptive kernel and
identifies specific velocity components \citep{1993Pisani};
when the chosen kernel is Gaussian, DEDICA reduces to one of the GMM methods.

Among the methods using both the galaxy positions and redshifts, the
Dressler \& Shectman (DS) method \citep{1988DS,1999Solanes,2000Knebe,2010Aguerri,Dressler2013} is certainly the
most widely used. 
Other methods include the three-dimensional (3D) KMM algorithm \citep{1994Bird,1996Colless,1998Barmby},
which, in input, requires the number of the substructures and an initial guess
of their positions; the 3D wavelet
transforms \citep{1992Escalera,1997Gambera,1997Girardi,1999Pagliaro}; 
the 3D version of DEDICA \citep{1996Pisani, 2007Ramella}; 
and the hierarchical tree algorithm
\citep{1996Serna,2005Adami}. 

All of these methods mainly focus on the substructure detection.
However, an unambiguous association of galaxies to individual substructures, which enables the derivation of the 
substructure properties, like size, velocity dispersion, and mass, is not yet available. 
Here we explore the possibility that the caustic technique 
can contribute a step forward in this direction.

The caustic technique \citep{1997Diaferio, 1999Diaferio, Diaferio2009}
estimates the escape velocity of cluster galaxies from the cluster center to 
a few times the virial radius.
The technique name derives from the two curves in the redshift diagram
where the galaxy number density is expected to be infinite in the spherical collapse model 
\citep{1989Regos}. With this technique, we can estimate the mass
and gravitational potential profiles of galaxy clusters to radii that extend to the cluster infall
region \citep[see reviews in][]{Diaferio2009,Serra2011}. For the mass estimation,
the caustic technique only
assumes spherical symmetry and does not require the system to be in
dynamical equilibrium. It can be used both in the central and in the outer
regions of clusters, where other techniques cannot be applied. 

The first step of the caustic technique procedure is to arrange
the galaxies in a binary tree according to a {\it projected} 
pairwise galaxy binding energy. This step is similar to the procedure described in \citet{1996Serna}.
However, the caustic technique goes further and identifies 
a threshold that cuts the 
tree and identifies the cluster members. \citet{2013Serra} showed that this approach,
combined with the location of the caustics in the cluster redshift diagram, returns 
a list of cluster members within $R_{200}$
that is 96\% complete and only 2\% of the members are actually interlopers. Within the 
larger radius $3R_{200}$, where no other method is available, the completeness is 95\% and 
the contamination is 8\%. 

The same principle used to cut the binary tree and identify the cluster members provides
a second threshold that gives a list of cluster substructures.
So far, no systematic analysis of the properties of these substructures
has been performed. Here, we use $N$-body simulations to investigate the
power of the caustic technique to identify cluster substructures.

In Section \ref{sec:nbody}, we describe the
cosmological $N$-body simulation and the mock cluster redshift surveys
we use to test the caustic technique as a substructure detector.
We briefly review the caustic technique in Section \ref{sec:caustic}. 
We present our results in Section \ref{sec:results}, and we discuss them in Section \ref{sec:discussion}. 

\section{The Simulated Cluster Samples}
\label{sec:nbody}

We use the Coupled Dark Energy
Cosmological Simulations \citep{2012Baldi}. This is the largest set to date of
$N$-body simulations that model the interaction between the dark energy scalar field and the Cold
Dark Matter (CDM) fluid.  Here, however, we only consider the simulation of the 
standard $\Lambda$CDM model with
fiducial WMAP7 parameters. The simulated volume is a comoving
cube of 1 $h^{-1}$~Gpc on a side
($ h = H_0 / 100$ km s$^{-1}$ Mpc$^{-1}$ is the dimensionless Hubble constant), containing $ 1024^3 $ CDM
particles with a mass of $5.84 \times 10^{10}h^{-1} M_{\odot}  $
and the same number of baryonic particles with a mass of $1.17 \times 10^{10} h^{-1} M_{\odot}$.  
We only consider the dark matter particles: we assume
that in the real universe galaxies are unbiased tracers of the velocity
field of the dark matter particles. In fact, both $N$-body simulations \citep[e.g.,][]{2001Diaferio,2004Diemand,2004Gill,2005Gill}
and observations \citep[e.g.,][]{2008Rines} indicate that any velocity
bias between galaxies and dark matter is smaller than 10\%. 

Halos are identified with 
the Friends-of-Friends (FoF) algorithm \citep{1982FoF,1985Davis}, which links particles 
with distances less than the linking length $l_{\rm FoF}$ to form a group. 
We adopt the standard linking length $l_{\rm FoF} = 0.2 l_{\rm mean}$,
with $l_{\rm mean}$ as the mean interparticle separation,
corresponding to the overdensity at viralization $\rho/\rho_{\rm b} = 185$ \citep{1998Audit},
with $\rho_{\rm b}$ as the mean background density. In this procedure, 
the FoF halos are identified by using the CDM particles as primary tracers and 
then linking baryonic particles to the group of their closest CDM neighbor. 
The characteristic radius of the FoF halos, $R_{200}$, is the radius within which the average density 
(including both CDM and baryonic particles) is 200 times the critical density. The mass within
$R_{200}$ is $M_{200}$.

We consider two samples of 50 FoF halos, each at redshift $z=0$:
a massive sample (M15 hereafter), with $M_{200}$ ranging from $0.86 \times 10^{15} h^{-1} M_{\odot}$ 
to $3.4 \times 10^{15} h^{-1} M_{\odot}$, and median $1.1 \times 10^{15} h^{-1} M_{\odot}$; and a less
massive sample (M14 hereafter), with masses ranging from $0.95 \times 10^{14}h^{-1} M_{\odot}$ 
to $1.1 \times 10^{14} h^{-1} M_{\odot}$, and median $1.0 \times 10^{14} h^{-1} M_{\odot}$.

For each cluster, we compile three mock galaxy redshift catalogs. 
Each cluster is located at the center of the volume using the periodic boundary conditions
of the simulation box.
We assign the celestial coordinates   
$(\alpha,\delta)=(6^h,0^\circ)$ and a redshift distance $cz=36,000$~km~s$^{-1}$
to the cluster center.
Around the cluster, we consider a rectangular prism enclosing the volume, 
corresponding to a solid angle that at the cluster
distance covers a square area $12 h^{-1}$~Mpc wide. The volume
is centered at the cluster and it is $140 h^{-1}$~Mpc deep.
The resulting field of view (FOV) is $1^{\circ}.6 \times 1^{\circ}.6$.
For each cluster, we apply this procedure to three orthogonal directions.
Since the clusters are generally not spherically symmetric, for our statistical purposes 
we can consider these three mock catalogs as independent clusters.
We thus obtain 150 mock redshift catalogs for each of the samples (M15 and M14). 

The observational volumes we extract 
from the simulation typically contain $\sim 8 \times 10^4$ particles for the M15 sample
 and $\sim 5 \times 10^4$ particles for the M14 sample.
Realistic numbers of observable galaxies in these volumes are clearly much smaller. 
Therefore, we randomly sample the dark matter particles until we obtain a given number of particles $N_{3R}$ 
within $3R_{200}$. To explore the effect of galaxy sampling, we build catalogs 
with $N_{3R}=$~(100, 200, 300, 400, and 500). Additionally, we only retain particles within 
$\pm 4000$~km~s$^{-1}$ from the cluster center.
These mock galaxy redshift surveys of cluster regions are roughly comparable 
to recent large surveys
of clusters and their surroundings, such as CIRS \citep{Rines2006} and
HeCS \citep{Rines2013}. 

A different strategy to build mock redshift surveys could be to keep the number of particles 
in the FOV for both the M14 and M15 samples fixed. 
However, this procedure returns mock surveys where the substructures of the M14 
clusters are poorly sampled or, more often, not sampled at all, because as mentioned above the observational 
volumes of the M14 sample are on average 60\% ($5 \times 10^4/8 \times 10^4$) 
less populated than the M15 sample volumes.   
On the contrary, keeping $N_{3R}$ fixed guarantees that we properly sample the substructures and guarantees that
it is closer to the observational procedure of surveys dedicated to the study of the dynamics of clusters; 
these surveys tend to sample the volume more densely around the cluster, both on the sky and in redshift space. 
The very different final numbers of particles in the FOV of the M14 and M15 samples 
(Table \ref{table:fields}) reflects a real effect: identifying substructures in less massive clusters 
requires denser surveys because in these clusters the probability of measuring the redshift of a galaxy 
that does not belong to the cluster is larger.

For a given $N_{3R}$, the total numbers of particles $N$ 
in the FOV of a mock cluster depend on the cluster and its surrounding region. 
To investigate the effects of the fluctuations caused by random sampling, we repeat the procedure 10 times. 
Table \ref{table:fields} lists the medians and percentile ranges of the number of particles $N$ as a function of $N_{3R}$:
for example, 80\%  of the mock catalogs of the M15 sample with $N_{3R}=100$ have $N$ in the range of $185-325$. 
For our statistical purposes, these 10 random realizations of each individual line-of-sight
projection of a given cluster with a fixed $N_{3R}$ cannot be considered independent. Hereafter, 
we will only use these 10 realizations to quantify the fluctuations of the random sampling on 
 the cluster samples M15 and M14.
Finally, in the M14 FOV's $N$ is four to five times larger than in the M15 samples: 
because the M15 clusters
are an order of magnitude more massive than the M14 clusters, whereas the surrounding cosmic
volumes are comparable, in the M15 samples the random sampling reaches $N_{3R}$ more rapidly 
and the number of particles sampled in the surrounding region is proportionally smaller. 

\begin{table}
\caption{The number of particles $N$ in the FOV.}
\centering
\begin{tabular}{|c|lll|lll|}
\hline
\multirow{2}{*}{$N_{3R}$} &\multicolumn{3}{|c|}{M15 (1500 clusters)} & \multicolumn{3}{|c|}{M14 (1500 clusters)} \\

\cline{2-7}
   & 10\% & 50\% & 90\% & 10\% & 50\% & 90\% \\
\hline
100  & 185 & 241 & 325 & 672 & 996 & 1490 \\
200  & 369 & 481 & 644 & 1356 & 2014 & 2905 \\
300  & 553 & 718 & 969 & 2034 & 2988 & 4385 \\
400  & 739 & 962 & 1283 & 2704 & 4016 & 5829 \\
500  & 923 & 1195 & 1603 & 3394 & 5034 & 7258 \\
\hline
\end{tabular}

\label{table:fields}
\end{table}

To identify the substructures of the clusters in the simulations,
 we use the code {\small SUBFIND} \citep{2001Springel}, whose algorithm is based on 
the overdensity and the gravitational binding energy of the particles. 
More specifically, for each FoF halo detected by the FoF algorithm,
{\small SUBFIND identifies candidate substructures by sorting the particles of the FoF halo according to their local density and isolating local density maxima. This procedure provides 
substructures whose boundaries are determined by the first saddle point identified in the local density field around each density maximum. From each substructure, we finally remove all particles with positive total energy }\citep[see][for further details]{2012Baldi}. 

The mass of a substructure is always its total mass, namely the sum of the mass 
of the particles (both CDM and baryons) that are gravitationally bound to that substructure 
as identified by {\small SUBFIND}.
The first row of Table \ref{table:subs} lists the total number of substructures
with masses larger than $10^{13} h^{-1} M_\odot$ in our two samples M15 and M14. 
The $10^{13} h^{-1} M_\odot$ mass threshold is not arbitrary, but is a minimum 
substructure mass set by the number of luminous galaxies that can be detected in current typical surveys. 
In fact, a $10^{13} h^{-1} M_\odot$ substructure is expected to contain at most a handful of galaxies brighter 
than $L_*$. Hereafter, we will call these substructures 3D substructures.

By randomly sampling the dark matter particles, 
the number of members of a 3D substructure in the mock catalog 
can be substantially reduced or even vanish.  
We only consider 3D substructures that have at least 10 particles appearing in the
FOV.  

Table \ref{table:subs} lists the total number of clusters $N_{\rm cl}$ with at least one 3D substructure 
appearing in the FOV, the total number of 3D substructures in all the FOVs $N_{\rm sub}$,
and the ratio between these detectable substructures
and the total number of substructures ($N_{\rm sub} / N_{\rm tot}$);
the total number of substructures $N_{\rm tot}$ is listed in the first row of Table \ref{table:subs}.
As expected, the number of 3D substructures appearing in the FOV increases with increasing $N_{3R}$. 
We also list the standard deviations deriving from the ten random realizations. We see
that the random sampling has a moderate impact. In the M15 samples, the number
of clusters that do not show 3D substructures in the FOV is substantial:
if we consider the total members
of the cluster as the sum of the members of the 3D substructures and the members of the cluster core 
identified by {\small SUBFIND}, $(36\pm 8)$\% of the total members belong to the 3D substructures in the M14 samples,  
whereas this fraction is only $(11\pm 6)$\% in the M15 samples. 
In addition, as mentioned earlier, the M15 FOV's are 4 to 5 times less populated than the M14 fields. 
Therefore, random sampling makes 3D substructures in the M15 samples vanish  
more easily than in the M14 samples.

\begin{table}
\caption{Number of clusters with 3D substructures in the FOV and number of 3D substructures.}
\centering
\begin{tabular}{|c|ccc|ccc|}
\hline
 \multirow{2}{*}{$N_{3R}$}   &\multicolumn{3}{|c|}{M15 } & \multicolumn{3}{|c|}{M14 } \\

\cline{2-7}
 & $N_{\rm cl}$ & $N_{\rm sub}$  & Ratio(\%) & $N_{\rm cl}$ & $N_{\rm sub}$ & Ratio(\%)\\
\hline
 -   & 150 	& 594 \tablenotemark{a} & - & 150  & 282 \tablenotemark{a} & -  \\
100  & 15 $\pm$ 1 & 15 $\pm$ 2  & 2.5 & 146 $\pm$ 1 & 191 $\pm$ 2 & 67.7  \\
200  & 39 $\pm$ 3 & 40 $\pm$ 3 & 6.7 & 150 $\pm$ 1 & 254 $\pm$ 4  & 90.1 \\
300  & 61 $\pm$ 4 & 64 $\pm$ 4 & 10.8 & 150 $\pm$ 0 & 275 $\pm$ 3  & 97.5 \\
400  & 81 $\pm$ 4 & 94 $\pm$ 5 & 15.8 & 150 $\pm$ 0 & 282 $\pm$ 1  & 100 \\
500  & 99 $\pm$ 3 & 132 $\pm$ 6 & 22.2 & 150 $\pm$ 0 & 282 $\pm$ 0& 100 \\
\hline
\end{tabular}
\tablenotetext{1}{ $N_{\rm tot}$, total number of 3D substructures with 
masses larger than $10^{13} h^{-1} M_{\odot}$ . }
\label{table:subs}
\end{table}

\section{The Caustic Method}
\label{sec:caustic}

According to hierarchical clustering models, clusters of galaxies form by the
aggregation of smaller systems. The local gravitational potential 
plays a crucial role in determining the galaxy velocities in addition to 
the radial infall expected in the spherical collapse
model \citep{1997Diaferio}. On the redshift diagram of the line-of-sight
velocity $v$ of the galaxies in the cluster rest frame versus their
projected distance $r$ from the cluster center, the cluster 
members populate a trumpet-shaped region that is approximately symmetric around
the $r$ axis \citep{Kaiser1987,1989Regos,1993vanHaarlem}.  The caustics
define the boundaries of this region, whose amplitude ${\cal A}(r)$
decreases with increasing $r$.  
${\cal A}(r)$ provides the estimate of the escape velocity profile
from the cluster and thus its mass profile \citep{1997Diaferio, 1999Diaferio}. 

To measure ${\cal A}(r)$, the caustic technique builds a binary tree based on the projected galaxy
pairwise energy, determines a threshold to cut the binary tree, identifies a set of candidate
cluster members that in turn determines the cluster center and defines 
the redshift diagram. The caustic technique locates the caustics and thus ${\cal A}(r)$
from the galaxy number density on the redshift
diagram. The steps that are relevant for the identification of the
substructures we are interested in here are the construction of the binary tree and
its threshold determination. 
For the sake of completeness, we list the details of these steps.
Further details are provided in \citet{1999Diaferio} and \citet{Serra2011}.

To build the binary tree, we proceed as follows: 
\begin{itemize}
 \item[i.] initially each galaxy is a group $g_\alpha$;
 \item[ii.] the binding energy $E_{\alpha\beta}={\rm min}\{E_{ij}\}$, where $E_{ij}$ 
is a projected binding energy between the galaxy $i\in g_\alpha$ and the
galaxy $j\in g_\beta$, is associated to each group pair $g_\alpha, g_\beta$. 
The projected binding energy is estimated with the relation 
\begin{equation}
E_{ij}=-G{m_i m_j\over R_{\rm p}}+{1\over 2}{m_i m_j\over m_i+m_j}\Pi^2 \;,
\label{eq:pairwise-energy}
\end{equation}
where $R_{\rm p}$ is the pair projected separation, $\Pi$ is the
line-of-sight velocity difference and $m_i=m_j=10^{12}h^{-1}$ 
M$_\odot$ is the typical total mass of a luminous galaxy; 

\item[iii.] the two groups with the smallest binding energy $E_{\alpha\beta}$ are
replaced with a single group $g_\gamma$ and the total number of groups is decreased by one;
\item[iv.] the procedure is repeated from step (ii) until only one group is left.

\end{itemize}

\begin{figure*}
\includegraphics[angle=0,height=0.72\textwidth,width=.96\textwidth]{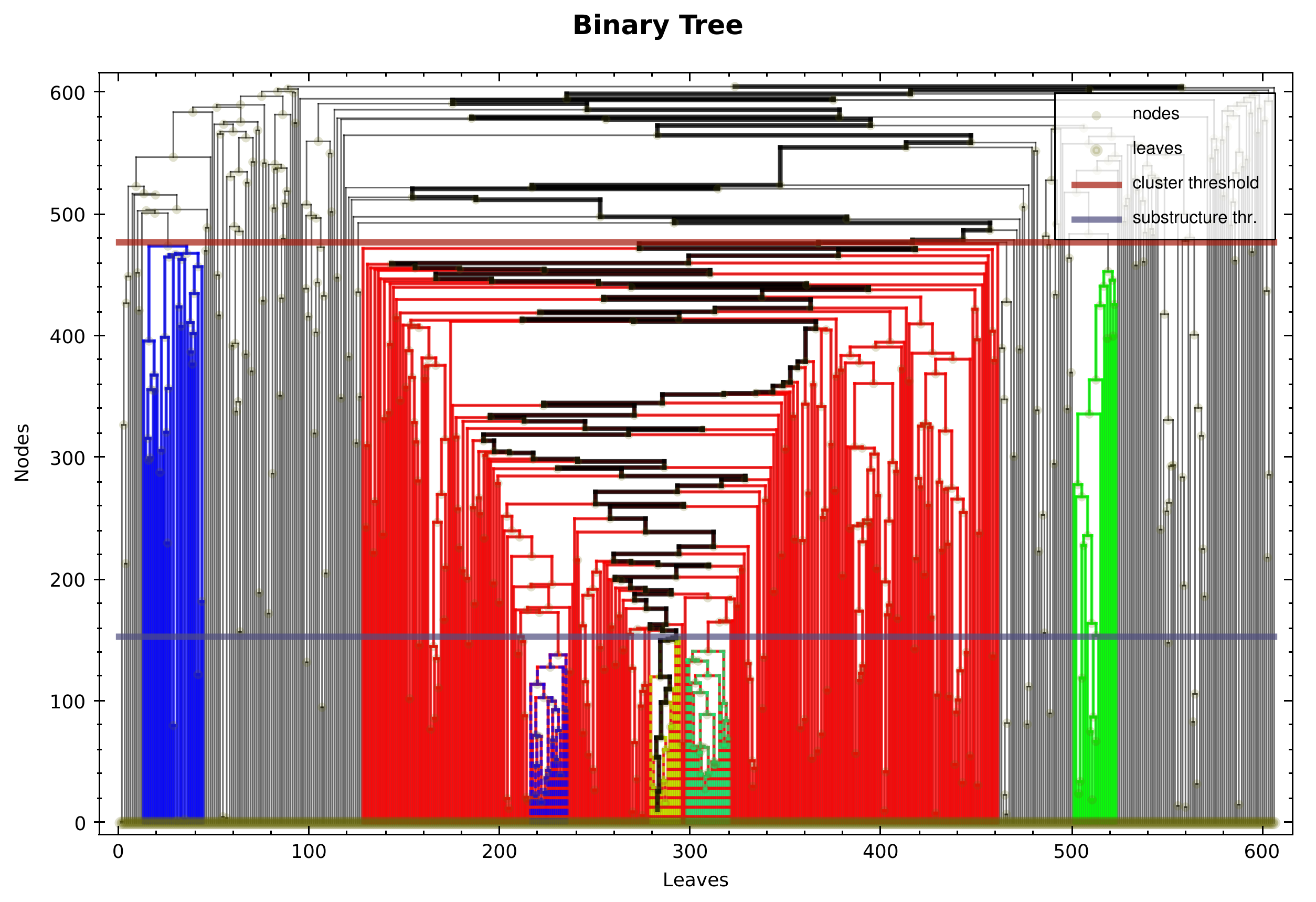}
\caption{Dendrogram of the binary tree of a simulated cluster from 
the M14 samples with 606 particles in the FOV. 
The particles are the leaves of the tree at the bottom of the plot. The thick path
highlights the main branch of the tree. The horizontal lines show
the levels at the two nodes $x_1$ (upper line) and $x_2$ (lower line)
that limit the $\sigma$ plateau shown in Figure \ref{fig:plateau}. The upper
node $x_1$ is the threshold that identifies the main group and the surrounding groups, 
while the lower node $x_2$ separates
the cluster substructures. The groups and substructures separated by the two thresholds 
are depicted with different colors: the main group is in red, 
two additional groups are in blue and green, the recognized core is in yellow, 
and the two substructures are depicted as blue and green dashed lines. This figure was 
generated with the software CausticApp (Serra and Diaferio, personalpages.to.infn.it/$\sim$serra/causticapp.html).} 
\label{fig:dendrogram}
\end{figure*}

\begin{figure}
\includegraphics[angle=0,width=0.48\textwidth]{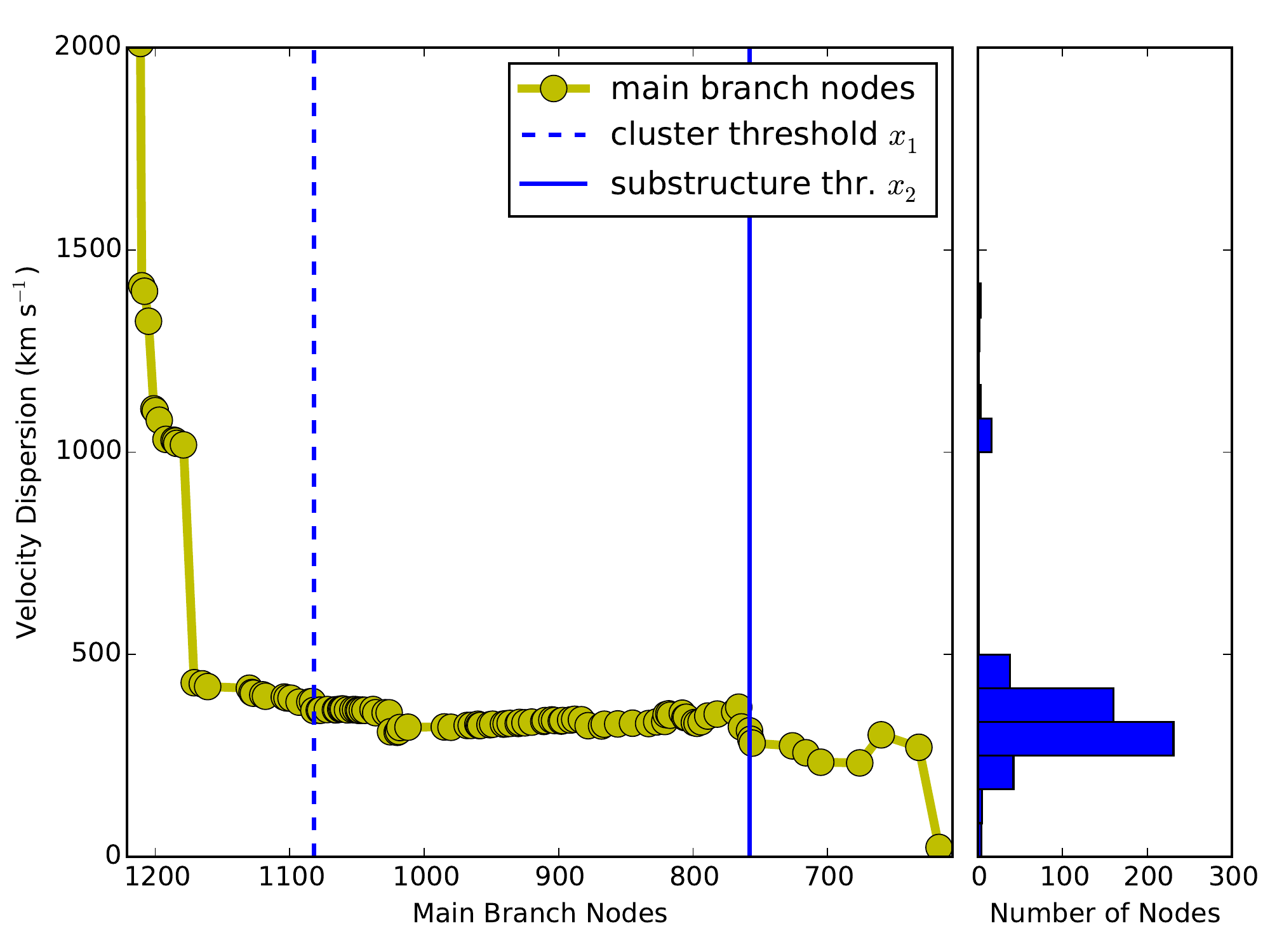}
\caption{Velocity dispersion of the leaves of each node
along the main branch of the binary tree shown in Figure \ref{fig:dendrogram}. 
The vertical dashed and solid lines show the nodes $x_1$ and $x_2$ respectively. 
The curve between $x_1$ and $x_2$ is
the $\sigma$ plateau, whose position is indicated by the peak of the histogram of
node numbers shown in the right panel.}
\label{fig:plateau}
\end{figure}

At this stage all the galaxies are arranged in a binary tree; an example is shown in Figure \ref{fig:dendrogram}.
This dendrogram is derived from a mock catalog of the M14 sample with $N_{3R}=100$ and $N=606$.
To identify the cluster members and its substructures, we need to cut the
tree at some level. Toward this aim, we identify
the main branch from the root to the leaves by tracing the node that contains 
the largest number of galaxies at each bifurcation.
 The leaves that hang from each node $x$ of the main
branch provide a velocity dispersion $\sigma_{\rm{los}}^x$.  When
walking along the main branch from the root to the leaves,
$\sigma_{\rm{los}}^x$ rapidly decreases due to the progressive loss of
galaxies that are most likely not associated with the cluster (Figure
\ref{fig:plateau}); then $\sigma_{\rm los}^x$ reaches a ``$\sigma$
plateau'' at some node $x_1$. Most of the galaxies hanging from this
node are members of the cluster: in fact, the system is nearly isothermal and moving
along the main branch toward the leaves removes the
less bound galaxies that in general do not substantially affect the value of
$\sigma_{\rm{los}}^x$.  When we get close to the leaves along the main branch,
the remaining galaxies have a binding energy that is 
very small and causes $\sigma_{\rm{los}}^x$ to drop again. 
This second rapid drop identifies the node $x_2$ that sets the limit of the
$\sigma$ plateau.
  
To identify the $\sigma$ plateau and its boundaries $x_1$ and $x_2$, \citet{Serra2011} 
designed an algorithm based on the distribution 
of the velocity dispersions of the nodes, as detailed below.
\begin{itemize}
\item[i.] Derive the probability density distribution of the velocity dispersion $\sigma_{\rm{los}}^x$ 
of the leaves hanging from each node; an example is 
shown in the right panel of Figure \ref{fig:plateau}.  
 The mode of this distribution corresponds to the value $\sigma_{\rm{pl}}$ of the $\sigma$ plateau. 
 \item[ii.] To identify the nodes belonging to the $\sigma$ plateau, (1) we remove
the tails beyond $\pm 0.3 \sigma_{\rm{pl}} $ of the $\sigma_{\rm{los}}^x$ distribution, 
and (2) the 80\% of the remaining nodes closest to $\sigma_{pl}$ are retained 
as the $N_{\delta}$ nodes defining the $\sigma$ plateau.
\item[iii.] We choose $x_1$ among the first (i.e., closest to the root) 
five nodes of the set of the $N_{\delta}$ nodes, 
as the node whose $\sigma_{\rm{los}}^x$ 
has the smallest discrepancy from $\sigma_{\rm{pl}}$; similarly we choose $x_2$ among the last five nodes 
(i.e., furthest away from the root).
\end{itemize}

The first node, $x_1$, closest to the root, is the
appropriate level for the identification of the cluster.
The threshold set by node $x_1$ separates the binary tree branches into 
different groups. The group containing the main branch is the main group
and its galaxies are the candidate members of the cluster. 
The completeness and purity of these candidate members have been 
investigated by \citet{2013Serra}. 
We consider all of the other groups separated by the threshold $x_1$ 
dynamically distinct from the cluster, or the main group, and 
we disregard them hereafter.

\begin{figure}
\includegraphics[angle=0,width=.48\textwidth]{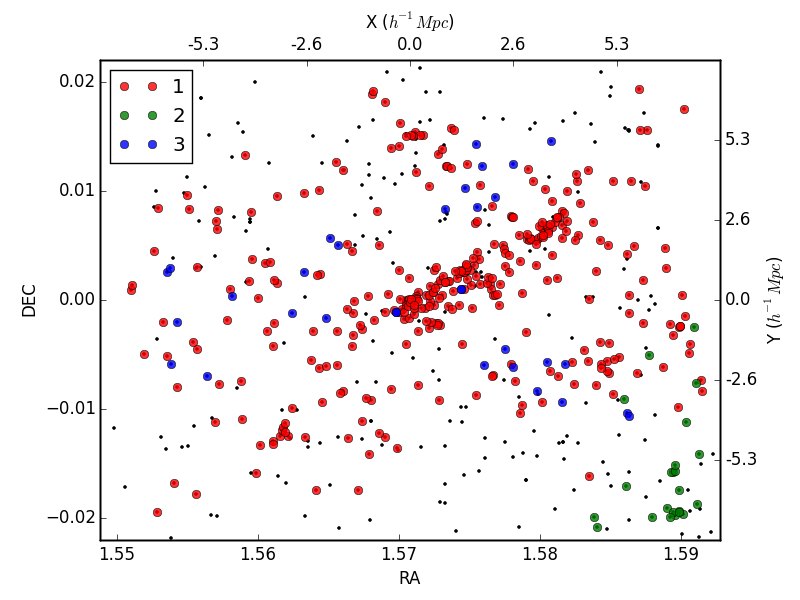}\\
\includegraphics[angle=0,width=.48\textwidth]{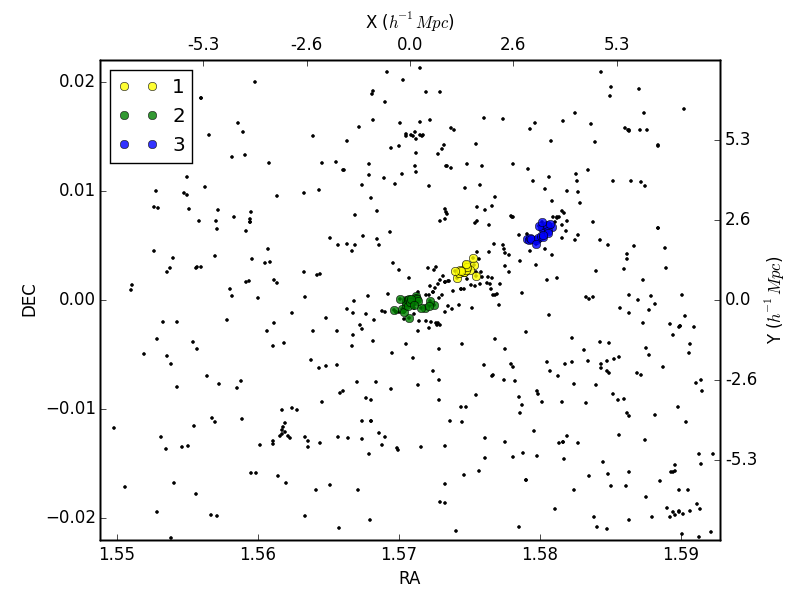}
\caption{Sky diagrams of the groups (upper panel) and 2D substructures (lower panel) of the
cluster whose dendrogram and $\sigma$ plateau are shown in Figures \ref{fig:dendrogram} and \ref{fig:plateau}. 
The projected celestial coordinates are in radiants; 
comoving coordinates in the $N$-body simulation are also shown.
Particles with the same color belong to the same structure identified by the binary tree: 
clusters and groups in the upper panel, 
and substructures in the lower panel. The color code is} the same as in Figure \ref{fig:dendrogram}.
\label{fig:diag_sky}
\end{figure}

The second node, $x_2$, farthest away from the root, identifies
the substructure candidates. We define all the substructures, whose members
belong to the main group, as the 2D substructures of the cluster.   
We only consider 2D substructures with at least 10 particles.
We disregard all of the systems separated by the threshold set by node $x_2$ whose members
do not belong to the main group. As an example, 
Figure \ref{fig:diag_sky} shows the distribution on the sky of the identified groups and the 2D substructures 
according to the dendrogram and $\sigma$ plateau  
of Figures \ref{fig:dendrogram} and \ref{fig:plateau}.

\section{Analysis and Results}
\label{sec:results}

\subsection{Cluster identification}

\begin{figure}[htbp]
\includegraphics[width=0.5\textwidth]{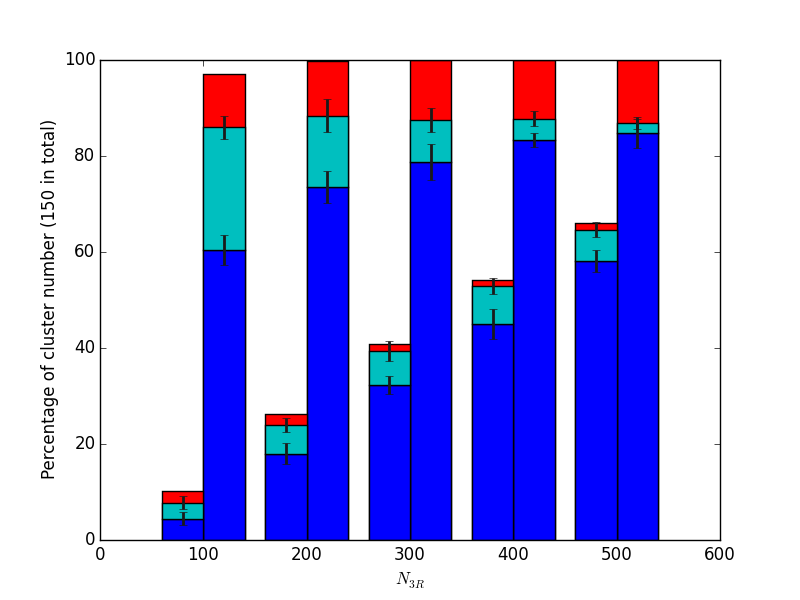}\\
\caption{Fraction of clusters with 3D substructures in the FOV. 
Around each $N_{3R}$, the left (right) bar is for the M15 (M14) sample.
The blue sectors show the fraction of correctly identified clusters 
with 2D substructures.
The cyan sectors show the fraction of correctly identified clusters without 2D substructures. The red sectors 
show the remaining fraction of clusters that are not on
the main branch of the binary tree. The error bars show the standard deviations
of these fractions deriving from the ten random sampling realizations.}
\label{clbar}
\end{figure}

Because of the random sampling of particles,
some substructures might have less than ten particles in the FOV.
According to our limit, these substructures are not taken into account. If all the 3D substructures
vanish from the field of view, the cluster has no substructures left.
The height of the bars of each $N_{3R}$ bin in Figure \ref{clbar} shows the total fraction
of clusters that have 3D substructures in the FOV. These fractions
correspond to the number of clusters listed in Table  \ref{table:subs}.
As noted in the previous section, almost all of the clusters in 
the M14 samples show 3D substructures, whereas many clusters
of the M15 samples have had all of their 3D substructures disappear.

When constructing the binary tree from the data set of a cluster redshift survey, 
the main group of the binary tree might identify a system different from the
cluster we are interested in because this cluster might not be the richest system in the FOV. 
With real data sets, where we usually analyze the clusters individually,
we can easily correct for this situation by reducing the area of the FOV or by imposing
the desired cluster center. Here, where we analyze large samples of mock clusters automatically and blindly,
we simply remove these cases. To check whether the main group identifies the correct cluster, 
we compare the 2D members, namely the members of the main group, with the
3D members of the cluster core identified by {\small SUBFIND}.  We say that a cluster is correctly identified if at least 60\% of its 3D members are in the list of the 2D members. 
In Figure \ref{clbar} the sum of the blue and cyan sectors of the bars shows the fraction
of correctly identified clusters; the red sectors of the bars show the fraction of misidentified clusters.

The fraction of  correctly identified clusters is further split into the fraction
of clusters with 2D substructures, shown by the blue sectors, and the fraction
of clusters with no 2D substructures, shown by the cyan sectors.
The lack of 2D substructures is usually caused by a relatively low second threshold
$x_2$ that does not leave a sufficiently large number of particles for the substructure identification.

In passing, we note that we confirm the results 
of \citet{2013Serra}, who find that on average in clusters identified 
by the $\sigma$ plateau, 13\% of the members within $3R_{200}$ are actually interlopers (see their Table 1, 
7$^{\rm th}$ column, 9$^{\rm th}$ row). Therefore, increasing our 60\% threshold only slightly decreases the 
fraction of correctly identified clusters, for example, only by 1\% if we increase the threshold  to 80\%. 
Substructures are more poorly populated and proportionally more difficult to detect than clusters. 
As we see below, for substructures, 
a 60\% threshold turns out to be a reasonable compromise between the completeness and the success rate: 
we thus also use the 60\% threshold for the clusters to adopt a single criterion for both structures. 

\subsection{2D versus 3D substructures: The success rate}

\begin{figure}[htbp]
\includegraphics[width=0.5\textwidth]{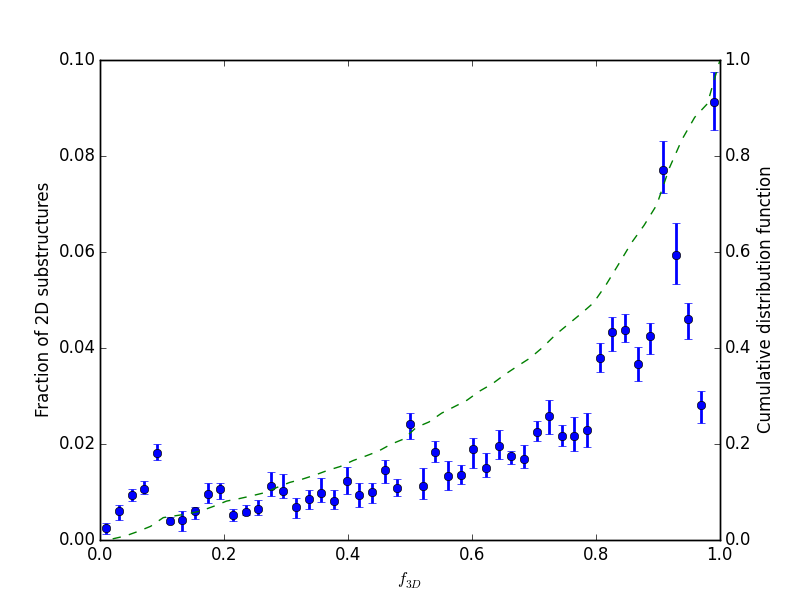}
\caption{Distribution of $f_{\rm 3D}$, the largest fraction of the total number of members of a 2D substructure
that are also members of a single 3D substructure.
The error bars show the 10\% and 90\% percentile ranges from the ten random sampling realizations.
The dashed line is the cumulative distribution function.}
\label{cmp_hist}
\end{figure}

To quantify whether the 2D substructures correspond to the 3D substructures, 
we make a one-to-one comparison 
between the members of the 2D substructures identified by the binary tree and the members of
the 3D substructures identified by {\small SUBFIND}. 
A single 2D substructure may contain members belonging to different 3D substructures or none. 
We find that in all of the M15 and M14 samples combined, 
49\% of the 2D substructures contain at
least one member of a 3D substructure. For each of these 2D substructures, 
we define $f_{\rm 3D}$ as the largest fraction of its total number of members
that are also members of a single 3D substructure.  
Figure \ref{cmp_hist} shows the distribution of $f_{\rm 3D}$: 51\% of the 
2D substructures have an $f_{\rm 3D}$ larger than 0.8. 
We adopt $f_{3D}=0.6$, a value smaller than the median $f_{3D}=0.8$, 
as the threshold to consider a 3D substructure
successfully identified by a 2D substructure. Adopting a smaller threshold 
increases the success rate at the expense of increasing the discrepancy
between the properties of the 2D and 3D substructures. A larger threshold
makes the identification more solid, but substantially drops the success rate. 

It can happen that different 2D substructures contain members of the same 3D substructure.
This event occurs for 12\% of the 2D substructures of the M14 samples and 
for 1.2\% of the 2D substructures of the M15 samples.
In these cases, we take the 2D substructure containing the
largest number of the 3D substructure members as the match to the 3D substructure. 

\begin{figure}[htbp]
\includegraphics[width=0.5\textwidth]{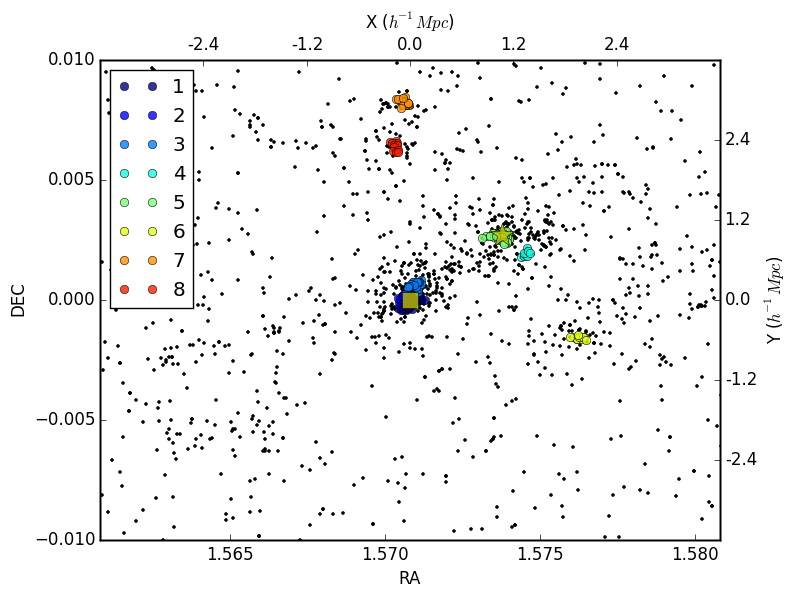}\\
\caption{Example of the identification of substructures with the caustic technique. 
This cluster from the M14 sample contains
500 particles within $3R_{200}$ with  4021 dark matter particles in the FOV (black dots).
Here we only show a fourth of the original FOV area around the cluster center indicated by the square.
The star shows the center of the only 3D substructure
with mass larger than $10^{13} h^{-1} M_{\odot}$.
The colored circles show the members of the seven 2D substructures and of the cluster core 
identified with the caustic technique.}
\label{prj}
\end{figure}

Figure \ref{prj} shows a random example of the substructure identification.
The cluster has only one 3D substructure, whose center is indicated by the yellow star. 
The caustic method returns seven 2D substructures in addition to the
cluster core, indicated by the yellow square, which is correctly matched.
One of the 2D substructures correctly coincides with the 3D substructure. Out of the remaining 
six 2D substructures that do not correspond to any 3D substructure of the cluster identified with {\small SUBFIND}, 
two are close to the core and four are relatively distant from 
the cluster center. We consider these six 2D substructures to be false detections.

\begin{figure}[htbp]
\includegraphics[width=0.5\textwidth]{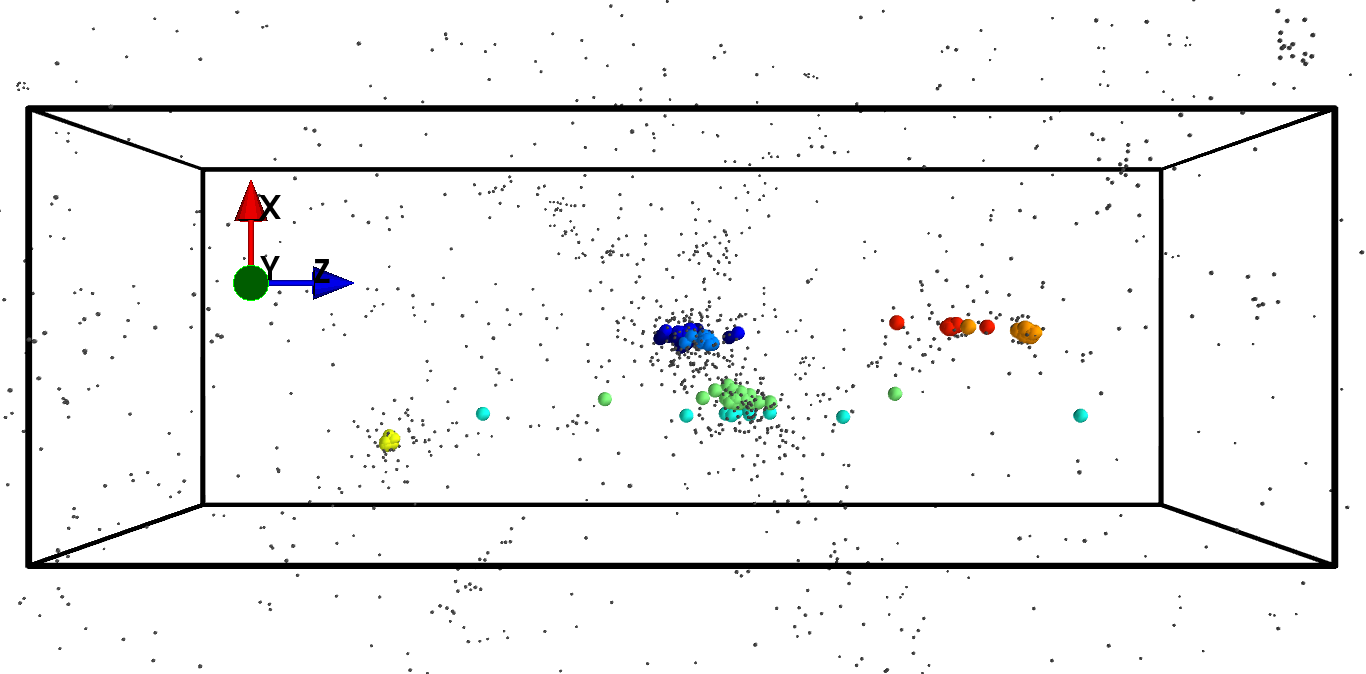}\\
\caption{The 3D distribution of the system shown in Figure \ref{prj}. To provide
the scale and a 3D perspective, we overplot a box with
dimensions $7 \times 7 \times 20 h^{-3}$~Mpc$^3$. The symbols
and colors are as in Figure \ref{prj}.}
\label{fig:3d4865}
\end{figure}

In fact, in our statistical analysis, we consider as false detections all the 2D substructures 
that do not correspond to the 3D substructures that  {\small SUBFIND} associates
with the analyzed cluster. However, our choice is rather restrictive.
Figure \ref{fig:3d4865} shows the 3D distribution of the system shown in Figure \ref{prj}: 
out of the six 2D substructures, only one system is due to chance alignment; the remaining
five 2D substructures are clearly bound systems. Two of them, close to the cluster core, 
are not in our list of 3D substructures because they have masses smaller than $10^{13} h^{-1} M_\odot$. 
The remaining three 2D substructures are groups surrounding the cluster
center. As mentioned above, we do not include them in our list of 3D substructures  in order 
to restrict our analysis to the 3D substructures of the analyzed cluster.

This random example shows that the rate of successful detections and the completeness we will show below 
are clearly correct in the context of focusing on the massive substructures of individual
clusters, but are likely to be lower limits to the performance of the 
identification of bound systems from redshift data with the caustic technique.

\begin{figure}[htbp]
\includegraphics[width=0.5\textwidth]{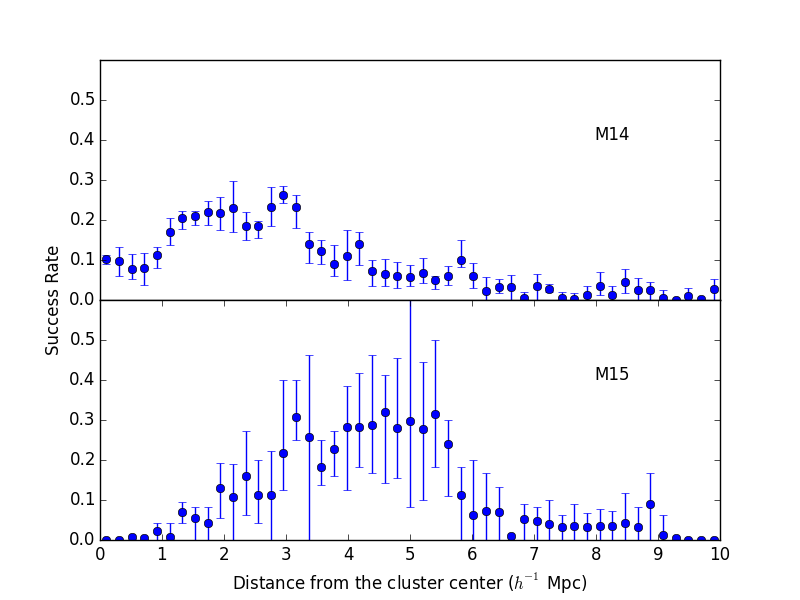}
\caption{Success rate as a function of the projected distance of the 2D substructures 
from the cluster center. 
The error bars show the 10\% and 90\% percentile ranges from the ten random sampling realizations.}
\label{dis_hist}
\end{figure}

Figures \ref{prj} and \ref{fig:3d4865} suggest that the 2D substructures at large projected
distances from the cluster center are likely to be groups and substructures of surrounding clusters. 
We thus compute the success rate, namely the ratio between the number of 2D substructures that 
correspond to 3D substructures and the total number of 2D substructures, as a function of
distance from the cluster center.  Figure \ref{dis_hist} shows this relation
for the combined M14 and M15 samples. 
The success rate peaks at different radii in the two samples because the
clusters have different sizes. When the radii are normalized to $R_{200}$, both 
peaks appear in the range $\sim 2-3R_{200}$.
We keep the length in this plot in proper units,  because these units are more convenient with real data; 
in addition, the mass distribution of each cluster sample is very peaked and 
the distance normalization plays a little role within the same cluster sample.
The success rate shows a broad peak between $\sim 1$ and $\sim 3 h^{-1}$~Mpc for the M14 samples and
between $\sim 3$ and $\sim 5 h^{-1}$~Mpc for the M15 samples, whereas it decreases
at the center and in the outskirts of the clusters.
The low rate at small distances is due to the cases where
the cluster core is identified as a substructure rather than as the core itself. 
Again, we consider these cases to be false detections because we are interested in the 3D substructure
identification, although these identified 2D substructures actually are bound systems.
In the cluster outskirts, the number of 3D substructures clearly decreases,
 unlike the number of 2D substructures.
Therefore, we can introduce a distance criterion by removing all of the 2D substructures at distances
outside a given range that, based on Figure \ref{dis_hist}, we arbitrarily
choose to be $[0.1, 6]h^{-1}$~Mpc: this criterion can remove 
most of the many false detections without 
missing promising 2D substructures. 

\begin{figure}[htbp]
\includegraphics[width=0.5\textwidth]{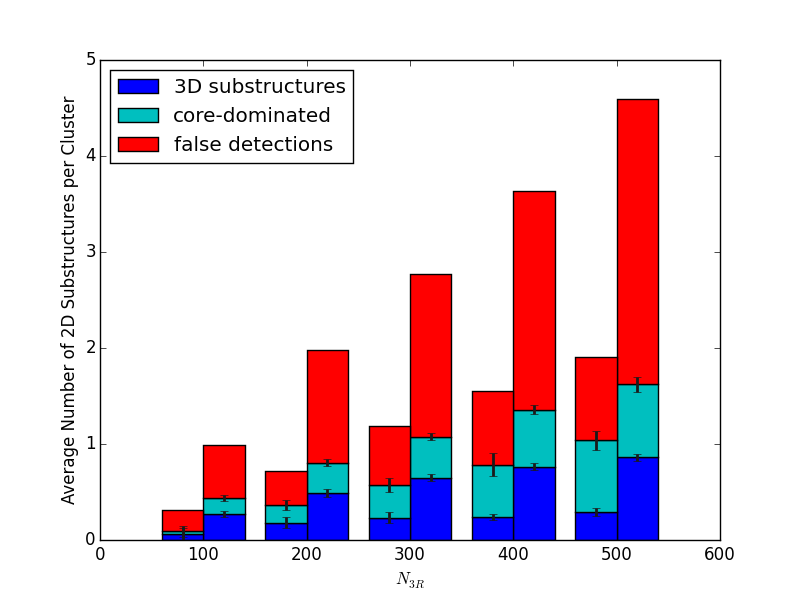}
\caption{Average number of 2D substructures per cluster as a function of $N_{3R}$: 
The left (right) bars are for the M15 (M14) samples. The  blue
sectors show the 2D substructures that correspond to 3D substructures;
the cyan sectors show the 2D substructures that are actually part of the cluster core; 
the red sectors show the false detections.
The error bars show the 1$\sigma$ deviation from the 10 random sampling realizations.}
\label{candi}
\end{figure}

Figure \ref{candi} shows the success rate as a function of $N_{3R}$, with the distance criterion applied. 
The average number of 2D substructures increases with $N_{3R}$, as expected, whereas
the success rate decreases from 25\% at $N_{3R}=100$ to 15\% at $N_{3R}=500$ (these fractions
can be read off in Figure \ref{candi} from the ratio between the length of the blue sector of each bar
and the total length of the bar): In fact, a larger number of particles
in the FOV increases the sampling of the 3D substructures, but at the same time it 
increases the probability of detecting 2D substructures by chance alignment.
The cyan sectors of the bars show the fraction of 2D substructures whose members are 3D members
of the cluster core rather than  members of the 3D substructures.
The error bars show that the random sampling fluctuations have a small impact.

\begin{figure}[htbp]
\includegraphics[width=0.5\textwidth]{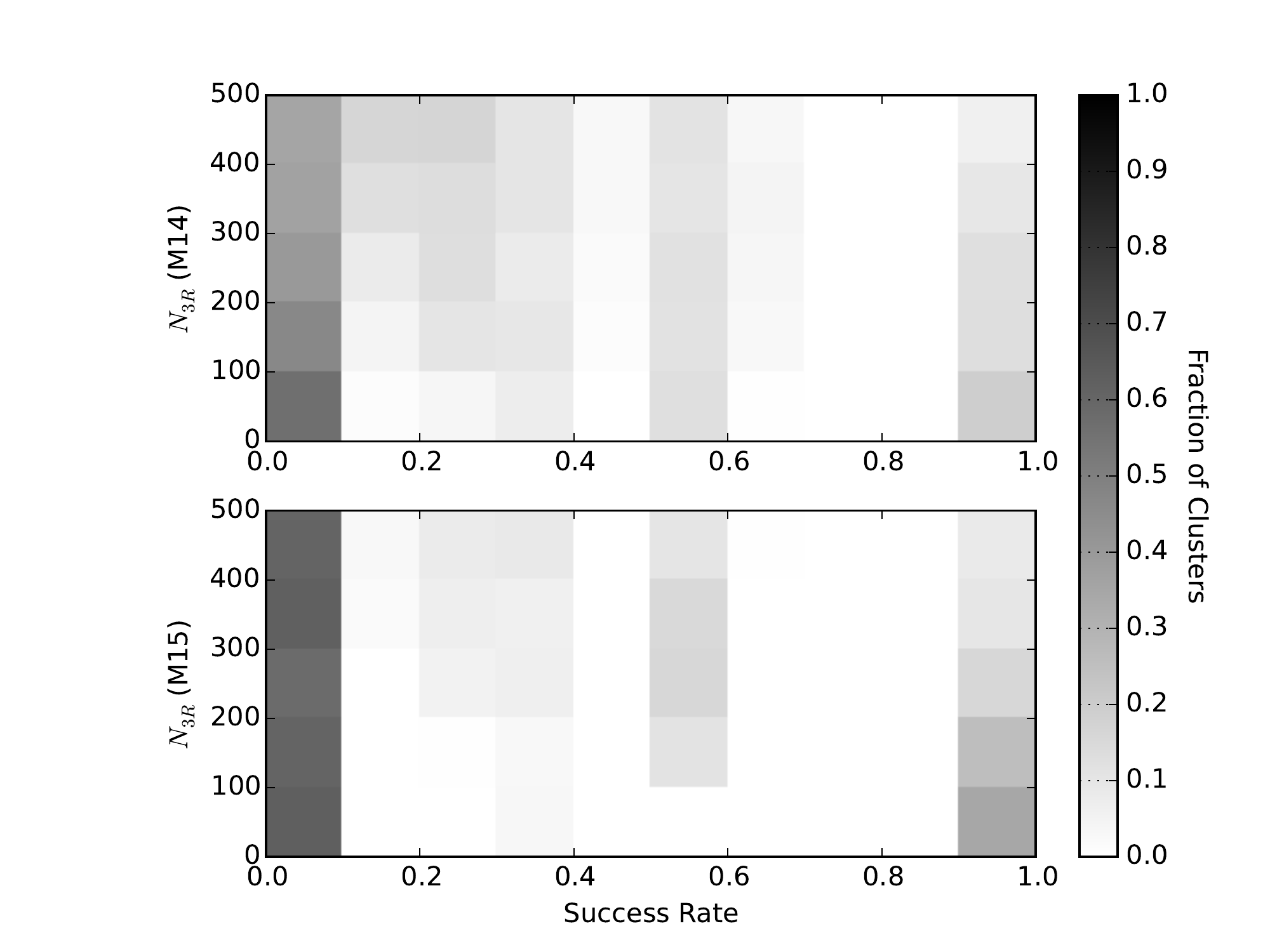}
\caption{The distribution of the clusters in the plane $N_{3R}$ vs. success rate.
The grey scale shows the number of clusters normalized to the total number of clusters 
with 3D substructures in the FOV.}
\label{suhist}
\end{figure}

We can look at our results from a different perspective.
Figure \ref{suhist} shows the distribution of the success rate. 
The y- and x-axes show $N_{3R}$ and the success rate, respectively.
The gray scale shows the fraction of clusters on this plane.
For example, the bottom row of the bottom panel shows that 63\% of the 
clusters in the M15 sample with $N_{3R}=100$ have no successful detection, whereas 
in 34\% of the clusters all the 2D substructures correspond to real 3D substructures.

This figure clearly shows that clusters are not uniformly distributed
on this plane. In fact, the number of 2D substructures is discrete and may be small,
with just one or two 2D substructures, especially when $N_{3R}$ is small.
Figure \ref{suhist} also shows 
that in all of the M15 samples, about 60\% of the clusters have 2D substructures that are false. 
This result is due to the fact that there are fewer particles belonging to 3D substructures in the FOV 
of massive clusters (see Sect. \ref{sec:nbody}).
The lack of substructures makes the boundary of the $\sigma$ plateau quite ambiguous
and, consequently the identification of the second threshold $x_2$ becomes more problematic.
On the contrary, the success rate in the M14 samples is more equally distributed on the plane and the
fraction of false 2D substructures is proportionally smaller.

\subsection{Completeness}

We now estimate the completeness of our samples of substructures, namely
the ratio between the number of correctly identified 3D substructures
and the total number of 3D substructures. We note that the total number of 3D substructures 
only includes the 3D substructures more massive than $10^{13} h^{-1} M_{\odot}$ 
and with at least 10 particles in the FOV.

\begin{figure}[htbp]
\includegraphics[width=0.48\textwidth]{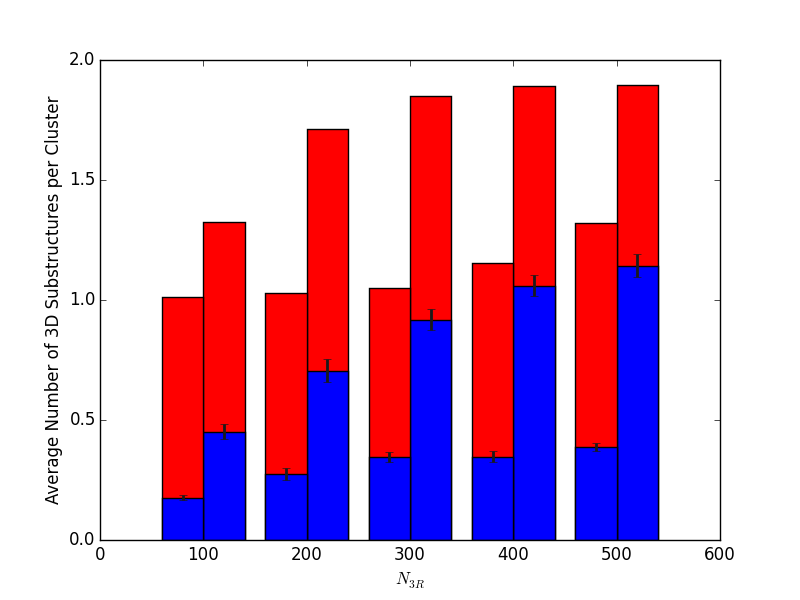} 
\caption{Completeness of the substructure catalogs. The left (right) bars are for the M15 (M14) samples. 
The blue sectors show the average number of 3D substructures properly 
detected in each cluster; the red sectors show the average number of  missed 3D substructures in each cluster.
The error bars  show the 1$\sigma$ fluctuations from the ten random sampling realizations.}
\label{fig:completeness}
\end{figure}

The completeness of the individual samples is shown in Figure \ref{fig:completeness}.
The completeness increases with $N_{3R}$ from  18\% to 29\% for the M15 samples, 
and from 34\% to 60\% for the M14 samples.
As for Figure \ref{candi}, these fractions
can be read off in Figure \ref{fig:completeness} from the ratio between the length of the blue sector of each bar
and the total length of the bar.

\begin{figure}[htbp]
\includegraphics[width=0.5\textwidth]{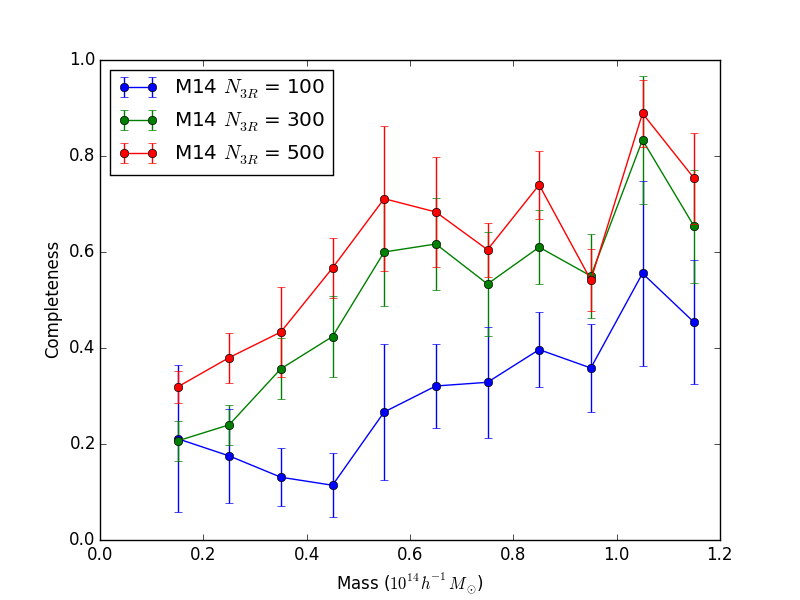}
\caption{Completeness vs. 3D substructure mass in the M14 samples for three different $N_{3R}$. 
The error bars show the 1$\sigma$ fluctuations from the ten random sampling realizations. }
\label{fig:m14hist}
\end{figure}

Figure \ref{fig:m14hist} shows the completeness as a function of the substructure mass
for the M14 samples.
Clearly, the most massive substructures are detected more easily, 
and more substructures are recognized in denser fields. 
The largest 3D substructure mass
is $1.19\times 10^{14} h^{-1} M_{\odot}$; this mass is larger
than the maximum $M_{200}$ of the M14 sample, but it still is a factor of $2.2$ smaller than 
the minimum total cluster mass
$2.66\times 10^{14} h^{-1} M_{\odot}$, which is defined as the sum of the mass of the dark
matter and baryonic particles of the FoF halo, consistently with the computation of
the mass of the 3D substructures.  
Figure \ref{fig:m14hist} shows that we can obviously improve the completeness of our substructure sample 
by increasing the lower mass limit. 
The result is qualitatively similar to the results of the M15 samples, which we do not show here, although
in this case, the trend has larger fluctuations and discontinuities due to the 
limited number of clusters and detected substructures. 

\begin{figure}[htbp]
\includegraphics[width=0.5\textwidth]{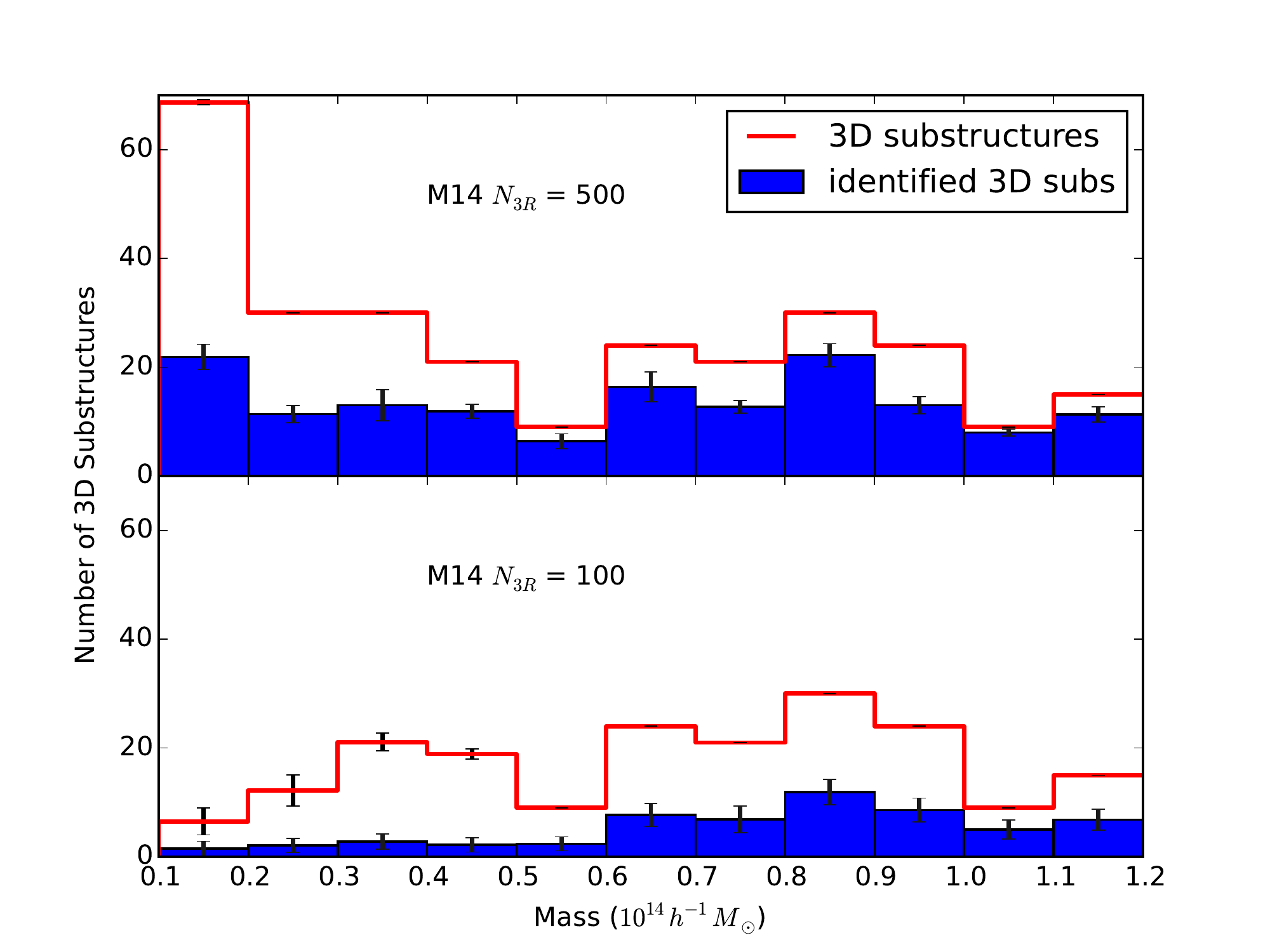}
\caption{Substructure mass functions.  
The upper (lower) panel is from the M14 sample with $N_{3R} = 500$ ($100$).
The open histograms show the real 3D substructure mass
function, whereas the solid histograms show the mass function of the detected 3D substructures.
The error bars show the 1$\sigma$ fluctuations from the ten random sampling realizations.}
\label{fig:mhist}
\end{figure}

To illustrate the origin of the relations plotted in Figure
\ref{fig:m14hist}, we show two examples of the substructure mass function in 
Figure \ref{fig:mhist}.
The red histogram is the original mass function of the 3D substructures in the
150-cluster M14 sample, whereas the blue histogram shows
the estimated mass function: the original mass function is not monotonic and is
severely under-represented by the estimated mass function at the low-mass end. 

This analysis of the dependence of the completeness on the minimum mass of
the 3D substructure unfortunately cannot be repeated with the success rate.
In fact, the substructure mass does not enter our algorithm for the identification of 
substructures from redshift data: changing the mass threshold would only change the 
number of 3D substructures, but it would leave
the number of 2D substructures unaltered. Therefore, if we increase the mass threshold, 
the success rate simply decreases; if we decrease the mass threshold below $10^{13}h^{-1} M_\odot$, 
we start probing substructures with one or two bright galaxies at most, namely substructures 
that are virtually impossible to detect with current redshift surveys.

\section{Discussion}
\label{sec:discussion}

\subsection{Possible developments of the caustic technique}

In the previous sections we show that the caustic technique, when blindly applied to redshift surveys of clusters, 
provides encouraging values of the success rate and completeness.
In principle, these results could be improved when a cluster is analyzed individually.
In fact, the assumption that the two nodes, $x_1$ and $x_2$, and their corresponding thresholds alone 
are sufficient to separate the groups surrounding the cluster as well as the cluster substructures 
might be too simplistic in some clusters: their dynamical state can be complex enough that the deviation from 
the single isothermal sphere is relevant. In this case, looking for a single $\sigma$ plateau on the
main branch might fail to capture
the full information on the cluster dynamics contained in the binary tree.

For example, the velocity dispersion of the main group can substantially differ 
from the velocity dispersion of the substructures, and in turn different substructures
can have different velocity dispersions. This situation can generate multiple $\sigma$ plateaus 
on the main branch and the determination of the proper thresholds is ambiguous.
In these cases, the algorithm might not identify the main group and its substructures satisfactorily
and might miss a substantial number of 3D substructures with velocity dispersions smaller than
the velocity dispersions set by the $\sigma$ plateau.

\begin{figure}[htbp]
\includegraphics[width=0.5\textwidth]{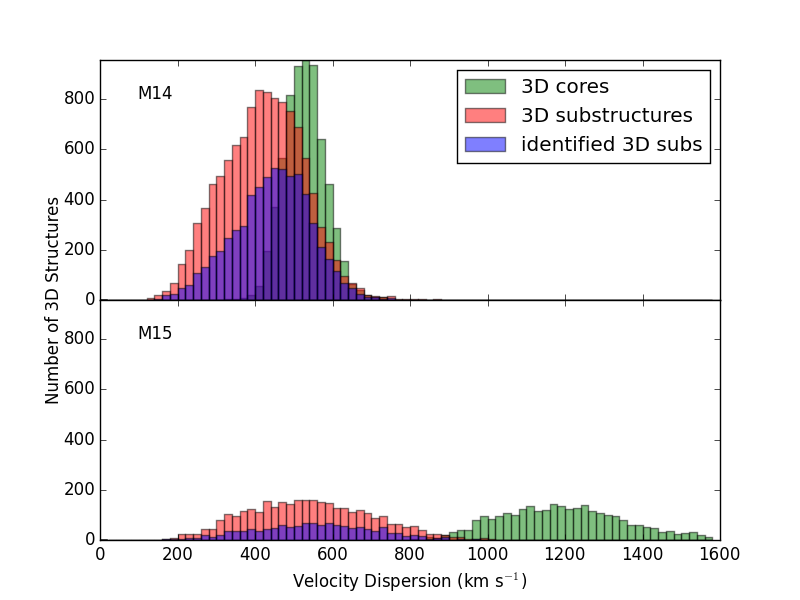}
\caption{Distributions of the velocity dispersions of the 3D substructures (red
histograms), 3D cluster cores (green histograms), and identified 3D substructures (blue histograms).
The upper (lower) panel shows the M14 (M15) cluster samples.
}
\label{fig:vdhist}
\end{figure}

Figure \ref{fig:vdhist} illustrates this argument: it shows the distributions of the velocity
dispersions of the 3D cores (green histograms), 3D substructures (red histograms), and the
2D substructures corresponding to 3D substructures (identified 3D substructures, blue histograms) 
for all of the combined M14 and M15 samples. 
The velocity dispersions of the 3D cores and substructures are computed with the full list of members
provided by {\small SUBFIND}, whereas the velocity dispersion of the identified 3D substructures are computed
from the list of members derived from the binary tree.
In the M14 samples, the velocity dispersions of the cores and of the 3D substructures
substantially overlap. Therefore the velocity dispersion  
corresponding to the single $\sigma$ plateau of the main branch of the binary tree
can properly identify both the main group and the substructures. 
On the contrary, in the M15 samples, the distributions of the velocity dispersions
of the cores and of the 3D substructures are almost completely separated and 
the $\sigma$ plateau that identifies the main branch is unlikely to alo properly
identify the substructures.  
Figure \ref{fig:vdhist} thus suggests that this is the origin of most of the differences 
between the substructure identification results of the M14 and M15 samples 
that we describe in the previous sections.

In conclusion, the closer the velocity dispersions of the 3D substructures are to the velocity dispersion 
of their cores, namely to the $\sigma$ plateau,
the more effectively our algorithm detects the 3D substructures. 
When applied to clusters individually, as it can happen with real catalogs containing 
fewer clusters, if the  shape  of the velocity dispersion on the main branch vs. the
node number (Figure \ref{fig:plateau}) is more complex than a single $\sigma$ plateau, 
the substructure detection can be improved by tuning the substructure threshold 
according to this shape. We plan to investigate this issue in detail in future work.

As a final note, we emphasize that our analysis is based on the assumption that 
a set of dark matter particles randomly sampled from an $N$-body simulation
is a fair representation of a real sample of galaxies. This assumption is expected
to be valid at the high-mass end of the dark matter halo mass function, but it can become
progressively incorrect at decreasing masses \citep{2014Sawala}.
Therefore, luminosity and velocity segregations can partly differentiate the
phase-space distributions of galaxies from those of an unbiased sample of dark
matter particles. Associating methods, like the abundance matching technique
\citep[SHAM;][]{2006Conroy,2006Vale},
which assumes a monotonic relationship between observed galaxy luminosity and simulated halo mass,
can be used to make tests on mock catalogs that resemble 
real data sets more closely. In future work we will explore this issue: we expect it to be particularly 
relevant in dealing with the different sizes of the M15 and M14 mock surveys (Table \ref{table:fields}) 
and in  addressing the dependence of the success rate on $N_{3R}$ (Figure \ref{candi})
in a more realistic context. 

\subsection{Our results in the context of previous work}

Assessing the presence of substructures in clusters 
has been frequently investigated in the literature based on X-ray observations,
weak gravitational lensing, or spectroscopic redshifts of galaxies. 
However, the issue has been addressed at very different levels.

X-ray analyses mainly focus on quantifying the surface brightness
morphology, with the aim of either separating relaxed clusters from merging clusters \citep[e.g.,][]{2015Parekh}
or quantifying the systematic errors
affecting cluster mass measurements \citep{2007Nagai,2008Piffaretti}.
The substructure identification is limited to the cluster central region
within $r_{500}$, which is the typical largest distance where the ICM can be 
detected reliably \citep{2008Piffaretti}, and to
substructures that contain a quantity of hot gas large enough to produce
a detectable X-ray emission.
In addition, the identification of substructures is
complicated by the fact that  viscosity and magnetic fields 
can displace the hot gas from the dominant mass distribution, 
as indicated by the observations of 
numerous merging clusters \citep[e.g.,][]{2004Markevitch,2007Mahdavi,2012Menanteau}.

The mass distribution is expected to be directly detected, with 
studies based on weak gravitational lensing, whose signal is not affected by the
complications of baryonic physics and the dynamical state of the cluster. 
However, unrelated large-scale structures along the line of sight \citep{2003Hoekstra, 2011Hoekstra}
and uncertainties on the lens parametrization  \citep{2004Clowe,2007Corless}
can be the sources of substantial systematic errors. For example, when weak lensing
is applied to the identification of clusters in $N$-body simulations, 
projection effects limit the fraction of identified clusters
that correspond to real clusters to 20\%, whereas 
the completeness can be as large as 80\% \citep{2002White}. 
\citet{2015Shirasaki} proposes a method  based on weak lensing to identify substructures in the 
outer regions of clusters, beyond 1~arcmin of the cluster center.  The method 
is tested on $N$-body simulations where
substructures are identified with {\small SUBFIND}. The fraction of real 
substructures that are identified with this algorithm is never larger than 40\%,
approximately comparable to our results, whereas the contamination fraction is not mentioned. 

The identification of substructures from the galaxy positions and redshifts has a long
history, but the various algorithms have rarely been tested on realistic $N$-body
simulations. \citet{1992Escalera} test the wavelet transform method on 12 toy models where 
the particles of the main cluster are distributed according to a 
Hubble profile, with a Gaussian distribution of their velocities;
only a single substructure, which is a rescaled model of the main cluster, is located
away from the cluster core. The wavelet transform method yields a significance
level of the identified substructure and \citet{1992Escalera} show
that for two of their significant substructures the fraction of particles that
do not actually belong to the 3D substructures is 3/9 and 0/7.

\citet{1996Pinkney} investigate 31 different methods, including the DS and the KMM algorithms;
only 5 out of these 31 methods use  both galaxy redshifts and celestial coordinates.
They build two-component merging clusters with $N$-body simulations
and apply the methods at different epochs of the merging process and from different lines of sight.
Therefore, the resulting mock fields of view contained a single substructure and no 
foreground  and background large-scale structures or field particles. 
The main goal of their test is to check the efficiency of the substructure
identification and the significance level of the detection.
The DS method has the best performance, with a success rate of 66\% atz  5\% significant level,
with a false detection rate of $\sim$ 10\%  in an isolated isothermal cluster.
The success rate of the KMM algorithm is around 36\%, while the false detection rate 
rapidly increases with the number of particles within the mock catalog. 

Systematic tests of the the DEDICA algorithm are performed by \citet{2007Ramella}.
They build artificial clusters with a single substructure away from the cluster center
and use Poisson noise to simulate the background and foreground structures.
The cluster and the substructure are spherically symmetric and have a King density
profile. On these toy models, most of the time DEDICA can detect the substructure 
and recover 75\% of its members. 

Unlike the studies described above, we use mock redshift surveys that are extracted 
from a state-of-the-art $N$-body simulation 
of a large cosmological volume, containing multiple substructures and 
foreground and background structures and filaments, thus
providing very realistic mock fields of view.
$N$-body simulations similar to ours were adopted by weak lensing analyses 
\citep{2002White,2015Shirasaki}.
To match our 2D substructures with the 3D substructures, we apply a
criterion based on the individual particles that are 
substructure members, a very strict but necessary criterion for assessing
the efficiency of identifying the 3D substructures from 2D information. 
Our strict criterion applied to realistic mock fields of view is the 
most relevant difference between our analysis and previous work. 
Therefore, comparing our results with the results of other techniques by looking
at their claimed performance alone can be misleading.
In this realistic context, the ability of our method to identify roughly $30$\% to $50$\% of
the genuine substructures of the cluster, independently
of its dynamical state and in the presence of coherent structures along the line of sight, is remarkable.
It also is important to emphasize that our substructure identification method only uses the first step
of the caustic technique, where the galaxies are arranged in a binary tree; this step does not
assume any spherical symmetry, unlike the unused second part of the caustic technique where the
redshift diagram and the caustic locations are determined. Therefore the substructure
identification technique does not assume any specific geometry for the substructure.

False identifications and incompleteness appear to be unavoidable in any technique. In future work, 
we plan to investigate some possible improvements, mentioned in 
the previous subsection, that are expected to reduce these failures and enhance
the performance of the caustic technique.

\section{Conclusions}
\label{sec:conclusion}

We test how efficiently the caustic technique can identify cluster substructures
in mock redshift surveys of clusters extracted from $N$-body simulations.
We consider two samples of 150 clusters each with $M_{200}\sim 10^{14}h^{-1}M_\odot$ (M14) and
$M_{200}\sim 10^{15}h^{-1}M_\odot$ (M15). 
We consider mock redshift surveys with different numbers of particles $N_{3R}$ within 
$3R_{200}$, including $N_{3R}=200$, which is the typical size of cluster redshift 
surveys like CIRS \citep{Rines2006} and HeCS \citep{Rines2013}.

For the $N_{3R}=200$ catalogs, 
among the 2D substructures identified by the caustic technique between $0.1$ and $6h^{-1}$~Mpc
from the cluster center, 17\% and 21\% 
correspond to the real 3D substructures with masses larger than $10^{13}h^{-1}M_\odot$ 
that are identified in three dimensions, for the M14 and M15 samples, respectively.
These numbers represent a lower limit to the numbers of physically bound
systems identified with the caustic technique because we also label as false detections 
real 3D substructures that are less massive than  $10^{13}h^{-1}M_\odot$ or that belong to groups
or clusters surrounding the cluster of interest.
As for the completeness, the lists of 2D substructures contain 48\% (M14) and 29\% (M15)
of the 3D substructures that are more massive than $10^{13} h^{-1} M_{\odot}$ and 
with more than 10 particles in the FOV.

Our analysis shows that the
completeness of the substructure catalog and the successful identification
of substructures is a strong function
of the substructure mass and the density of the survey. However, this latter parameter
does not necessarily need to be as large as possible, because
denser surveys have a larger probability of chance alignments and the association
of interlopers. Quantifying these effects more systematically requires
further investigation.

We show that the caustic method appears to be a promising technique
for identifying substructures of galaxy clusters out to their outer regions
from redshift surveys. When used for this purpose, 
the caustic technique does not require the assumption of spherical symmetry, 
and it is thus an ideal tool for analyzing complex systems.
However, the method can certainly be improved: 
the caustic method arranges the galaxies
in a binary tree based on a projected binding energy; the information on
the dynamical state of the cluster contained in this binary tree is impressively
rich and deserves further investigation to be fully exploited.

An efficient technique for investigating the substructure content of clusters 
is well-timed because data sets, including both redshift surveys  of
the large-scale structures, like
SDSS \citep{2014Ahn} and LAMOST \citep{2012Zhao}, and dedicated redshift surveys of clusters,
like CIRS \citep{Rines2006} and HeCS \citep{Rines2013}, are already available.

\acknowledgments

We sincerely thank Margaret Geller and Ken Rines for valuable suggestions
and an anonymous referee whose detailed comments greatly helped us to illustrate
our results more clearly. We 
acknowledge support from the grant Progetti di
Ateneo/CSP$\_$TO$\_$Call2$\_$2012$\_$0011 ``Marco Polo'' of the University of Torino, the INFN grant InDark, the grant
PRIN 2012 ``Fisica Astroparticellare Teorica'' of the Italian
Ministry of University and Research, the Ministry of Science and Technology
National Basic Science Program (Project 973)
under grants Nos. 2012CB821804, and 2014CB845806, the Strategic
Priority Research Program ``The Emergence of Cosmological Structure"
of the Chinese Academy of Sciences (No. XDB09000000), the National
Natural Science Foundation of China under grants Nos. 11373014,
11403002, and 11073005, the Fundamental Research Funds for the
Central Universities, and the Scientific Research Foundation of Beijing
Normal University.
M.B. also acknowledges the financial contribution by the PRIN INAF 2012 ``The Universe in the box: multiscale simulations of cosmic structure.''

\bibliography{subgr}

\end{document}